\documentclass[11pt]{article}
\usepackage{Blois,epsfig}

\def\be{\begin{equation}}
\def\ee{\end{equation}}
\def\bea{\begin{eqnarray}}
\def\eea{\end{eqnarray}}

\newcommand{\beq}{\begin{equation}}
\newcommand{\eq}{\end{equation}}

\begin{document}
\vspace*{2cm}

\vspace*{2cm}
\title{
 HARD PRODUCTION OF EXOTIC HYBRID MESONS }

\author{ I. ANIKIN$^{1,2,5}$, B. PIRE$^2$,
  L. SZYMANOWSKI$^{3,4}$, O.V. TERYAEV$^1$ and S. WALLON$^5 $}

\address{${}^1$\,Bogoliubov Laboratory of Theoretical Physics, JINR, 
141980 Dubna,
Russia \\[0.2\baselineskip]${}^2$\,CPHT$\;\,$\footnote{Unit{\'e} mixte C7644 
du CNRS}, \'Ecole
Polytechnique, 91128 Palaiseau, France \\[0.2\baselineskip]
${}^3$\,Soltan Institute for Nuclear Studies, Warsaw, Poland
\\[0.2\baselineskip]
${}^4$\,Universit\'e  de Li\`ege,  B4000  Li\`ege, 
Belgium\\[0.2\baselineskip]
${}^5$\,LPT$\;\,$\footnote{Unit{\'e} mixte 8627 du CNRS}, Universit\'e 
Paris-Sud, 91405-Orsay, France  \\
}

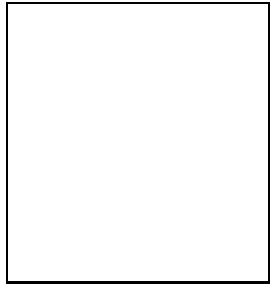
\maketitle\abstracts{
Exotic hybrid mesons $H$, with quantum numbers $J^{P\,C}= 1^{-\,+}$ may 
be copiously
produced in the hard exclusive processes
$\gamma^* (Q^2)\gamma \to H$  and $\gamma^* (Q^2) P(p) \to H P(p')$
because they have a leading twist distribution amplitude with a sizable
 coupling constant $f_H$,
which  may
be estimated through QCD sum rules. The reaction rates scale in the 
same way as the
corresponding rates for usual mesons. }

\section{ Introduction}

Within quantum chromodynamics, hadrons are described in terms of 
quarks, anti-quarks and
gluons. The usual, well-known, mesons are supposed to contain quarks and
anti-quarks as valence
degrees of freedom while gluons play the role of carrier of interaction.
On the other hand, QCD does not prohibit the existence of the explicit 
gluonic degree of freedom
in the form of a vibrating flux tube, for instance. The states where 
the $q\bar q g$ and
$gg$ configurations are dominating, hybrids and glueballs,
are of fundamental importance to understand the dynamics of quark 
confinement and the
nonperturbative sector of quantum chromodynamics.
We investigated \cite{APSTW} how hybrid mesons with $J^{PC}=1^{-+}$, 
one candidate  being $\pi_1(1400)$,
may be studied through the so-called hard reactions.
We focussed on deep exclusive   $\gamma^* P $ 
reactions
  which are well described in the framework of the collinear
approximation where generalized parton distributions (GPDs)~\cite{Muller:1994fv} and
(generalized) distribution amplitudes \cite{ERBL,GDA} describe the 
nonperturbative parts
of a factorized amplitude \cite{ERBL,Collins:1997fb}. We now suggest to use 
$\gamma^* \gamma $ reactions at $e^+\;e^-$
colliders to reveal such hybrid mesons.

\section{Hybrid meson distribution amplitude}

Let us consider the properties of the longitudinally polarized hybrid meson
distribution amplitude.
The Fourier transform of the hybrid meson--to--vacuum matrix element of
the bilocal vector quark operator may be written as
\begin{eqnarray}
\langle H_L(p,0)| \bar \psi(-z/2)\gamma_\mu
\psi(z/2) | 0 \rangle =
i f_H M_H e^{(0)}_{L\,\mu}
\int\limits_0^1 dy e^{i(\bar y - y)p\cdot z/2} \phi^{H}_L(y)
\label{hmeW},
\end{eqnarray}
where $e^{(0)}_{L\,\mu}=(e^{(0)}\cdot z)/(p\cdot z) p_\mu$
and $\bar y=1-y$ and $H$ denotes the isovector triplet of hybrid mesons;
$f_H$ denotes a dimensionful coupling constant of the
hybrid meson, so that $\phi^H$ is dimensionless.

In (\ref{hmeW}),
we imply the path-ordered gluonic exponential along the straight line 
connecting
the initial and final points $[z_1;z_2]$ which provides the gauge 
invariance
for bilocal operator and
equals unity in a light-like (axial) gauge. For  simplicity of notation 
we shall omit the index
$L$ from the  hybrid meson distribution amplitude.

Although exotic quantum numbers like $J^{PC} = 1^{-+}$ are forbidden in 
the quark model,
it does not prevent the
leading twist correlation function from being non zero. The basis of 
the argument is that the
non-locality of the quark correlator opens the possibility of getting 
such a hybrid
state, because of dynamical gluonic degrees of freedom arising from the 
Wilson line
(more details can be found in \cite{APSTW}).

To estimate the magnitude of the distribution amplitude, on uses the
estimate of the matrix element 
 of the quark energy-momentum tensor between 
vacuum and
exotic meson state which was explored long ago \cite{BD}.
It may be related, by making use
of the equations of motion, to the matrix element of quark-gluon 
operator
and  estimated with the help of the techniques of QCD sum rules.
One of the solutions corresponds to a resonance with  mass
around $1.4\, {\rm GeV}$ and the coupling constant
$ f_{H } \approx 50 \,{\rm MeV}$.

In summary, the hybrid light-cone distribution amplitude is a leading 
twist quantity
which should have a vanishing first moment because of its symmetry 
properties.
It obeys usual evolution equations \cite{ERBL}
and has an asymptotic limit
\begin{eqnarray}
\label{Has}
\Phi^H_{as} = 30  y (1-y) (1-2y)\;,
\end{eqnarray}
which allows to fix the value of the coupling constant $f_H$ assuming
normalization of the distribution amplitude $\Phi^H(y)$ as
$$
\int\limits_0^1 dy (1-2y)\Phi^H(y)=1\;.
$$

\section{$\gamma^* (Q^2)\gamma \to H$ }
The reaction $\gamma^* (Q^2)\gamma \to H$ may be studied at intense 
electron-positron
colliders such as CLEO, BABAR and BELLE. The production amplitude is 
proportional
to the hybrid meson distribution amplitude $\Phi_H(z)$. It reads


\begin{equation}
\label{ampl}
T_{\mu\nu}(\gamma\gamma^*\to H_L)=g_{\mu\nu}^T
\frac{(e_u^2 -e_d^2)f_H}{2\sqrt{2}}\;
\int\limits_0^1\,dz\,\Phi^H(z)\left(\frac{1}{\bar z} - 
\frac{1}{z}  \right)\,,
\end{equation}
with $g_{\mu\nu}^T$ projecting on the transverse coordinate subspace. 
It is convenient to study the cross-section ratio
\begin{equation}
\label{Hpi}
\frac{d\sigma^{H}}{d\sigma^{\pi^0}}=
\biggl|\frac{f_H\;\int\limits_0^1 \,dz\;\Phi^H(z)\left(\frac{1}{z}
-\frac{1}{\bar z}  \right)}{f_\pi
\;\int\limits_0^1 \,dz\;\Phi^\pi(z)\left(\frac{1}{z}
+\frac{1}{\bar z}  \right)
} \biggr|^2\;.
\end{equation}
We want to stress that
using asymptotic distribution amplitude  for the hybrid meson (\ref{Has})
and for the pion $\Phi^\pi(z)=6z\bar z$, the obtained  
ratio of cross sections is quite large,
$d\sigma^{H}/d\sigma^{\pi^0} \approx 38\%$. 
If the process of reference is taken as $\gamma^* \gamma \to \eta$ 
then the corresponding ratio is
$d\sigma^{H}/d\sigma^{\eta} \approx 46\%$.


One may have to base
the hybrid identification process through one of its possible decay
products. If the  $\pi\eta $ decay mode is  dominant
as seen for $\pi_{1} (1400)$  (or $\pi \eta'$ for
the candidate $\pi_{1} (1600)$),
we need to introduce  the concept of generalized distribution
amplitude (GDA) 
for $\pi \eta$. 
This factorized
description of exclusive processes in $\gamma^* \gamma$ collision has been
shown \cite{APT} to be successful for  $Q^2$ greater than a few GeV$^2$.
The $\pi^0\eta$ GDA may be
defined
  as :
\begin{equation}
\label{hme2}
\langle \pi^0(p_\pi)\eta(p_\eta) |
\bar\psi_{f_2}(-z/2)\gamma^{\mu}[-z/2;z/2] \tau^3_{f_{2}f_{1}}
\psi_{f_1}(-z)|0\rangle=
p^{\mu}_{\pi\eta}\int\limits_{0}^{1}dy e^{i(\bar y-y)p_{\pi\eta}\cdot
z/2}
\Phi^{(\pi\eta)}(y,\zeta, m_{\pi\eta}^2),
\end{equation}
where  the total momentum of $\pi\eta$ pair 
$p_{\pi\eta}=p_{\pi}+p_{\eta}$,
while $m^2_{\pi\eta}=p^2_{\pi\eta}$.
This $\pi\eta$ distribution amplitude $\Phi^{(\pi\eta)}$ describes non
resonant
as well as resonant contributions. The $\zeta$-parameter is defined
as :
\begin{equation}
\label{zeta}
\zeta=\frac{p_\pi^+}{(p_\pi+p_\eta)^+}-
\frac{m^2_\pi-m^2_\eta}{2m^2_{\pi\eta}},
\end{equation}
which is  related to the angle $\theta_{}$,
defined as the polar angle of the $\pi$ meson in the center of mass
frame of the meson pair, through:
\begin{eqnarray}
\label{2zeta}
2\zeta-1= \frac{\lambda(m^2_{\pi\eta},m^2_\eta,m^2_\pi)}{m^2_{\pi\eta}}
\; \cos\theta_{}.
\end{eqnarray}

Since the mass region around $1400$ ${\rm MeV}$ is dominated by the
strong $a_2(1329)\,(2^{++})$
resonance, it is interesting to look for the
interference of the amplitudes of hybrid and $a_{2}$ production, which
is linear, rather
than quadratic in the hybrid electroproduction amplitude.
We model the $\pi\eta$ distribution amplitude in the following form:
\begin{eqnarray}
\label{approx}
\Phi^{(\pi\eta)}(y,\zeta, m_{\pi\eta}^2)=30 y (1-y)(2y-1)
\biggl[
\,B_{11}(m_{\pi\eta}^2) P_1(\cos\theta) +
B_{12}(m_{\pi\eta}^2)
P_2(\cos\theta)
\biggr],
\end{eqnarray}
with the coefficient functions $B_{11}(m_{\pi\eta}^2)$ and
$B_{12}(m_{\pi\eta}^2)$   related to corresponding Breit-Wigner
amplitudes when
$m^2_{\pi\eta}$ is in the vicinity of $ M^2_{a_2},\,M^2_H$.

\section{Deep exclusive electroproduction}
The exotic hybrid meson may also be studied by means of its deep 
exclusive
electroproduction, {\it i.e.}
\begin{eqnarray}
\label{pr}
\gamma^*_L(q)\, + \, N(p_1)\,\to\, H_L(p)\,+\,N(p_2)
\end{eqnarray}
when the baryon is scattered at small angle and the transferred 
momentum $-q^2=Q^2$
is  large (Bjorken regime).
Within this regime, a factorization theorem is valid  at
the leading twist level, and a partonic subprocess part described in 
perturbative
QCD
is convoluted with universal soft parts, which are GPDs and meson
distribution amplitudes.
The leading order amplitude for the process (\ref{pr}) is
\begin{eqnarray}
\label{amp02}
{\cal A}_{(q)}= \frac{e\pi\alpha_s f_{H} C_F}{\sqrt{2}N_c Q}
\biggl[ e_u {\cal H}_{uu}^- -e_d {\cal H}_{dd}^-\biggr] {\cal 
V}^{(H,\,-)},
\end{eqnarray}
where
\begin{eqnarray}
\label{softin1}
&&\hspace*{-1cm}{\cal H}_{ff}^\pm =
\int\limits_{-1}^{1}dx \biggl[
\overline{U}(p_2)\hat n U(p_1) H_{ff^{\prime}}(x) +
\overline{U}(p_2)\frac{i\sigma_{\mu\alpha} 
n^{\mu}\Delta^{\alpha}}{2M}U(p_1)
E_{ff^{\prime}}(x) \biggr]
\biggl[
\frac{1}{x+\xi-i\epsilon}\pm\frac{1}{x-\xi+i\epsilon}\biggr],
\nonumber \\
&&
{\cal V}^{(M,\,\pm)}=
\int\limits_{0}^{1} dy \phi^{M}(y)\biggl[
\frac{1}{y}\pm\frac{1}{1-y}
\biggr].
\end{eqnarray}
Here, functions $H$ and $E$ are standard leading twist GPD's.
In (\ref{softin1}), we include the definition of ${\cal H}_{ff}^+$ and
${\cal V}^{(M,\,+)}$ which are useful for the comparison with the 
$\rho$ meson
case.
We discussed in details elsewhere \cite{APSTW} the resulting cross 
section. Let us just quote here the estimate for the ratio of hybrid
and $\rho$ meson electroproduction
cross-sections:
\begin{equation}
\label{ratio}
\frac{d\sigma^{H}(Q^2, x_B, t )}{d\sigma^{\rho}(Q^2, x_B, t )}=
\biggl|\frac{f_H}{f_\rho}
\frac{( e_u {\cal H}_{uu}^- -e_d {\cal H}_{dd}^-) {\cal V}^{H}}
{( e_u {\cal H}_{uu}^+ -e_d {\cal H}_{dd}^+) {\cal V}^{\rho}}\biggr|^2 \approx
13\% \;\;\;\;\;\mbox{for}\;\;\;\;x_B=0.33\;.
\end{equation}
For smaller $x_B$, e.g. for $x_B=0.18$, this ratio of cross sections is
roughly $3\%$.
\section{Conclusion}

In conclusion, we have calculated
the leading twist contribution to exotic hybrid meson with 
$J^{PC}=1^{-+}$ production
amplitude in the deep exclusive region.  The resulting order of 
magnitude
of the cross sections is
sizeable and should be measurable at dedicated experiments at BABAR and 
BELLE on
the one hand, and at JLab, Hermes or Compass on the other hand. We also 
estimated the
effects of next to leading order contributions \cite{APSTW3}.
In the region of  small $Q^2$  higher twist contributions should be 
carefully studied and
included. We  leave this study for future works.

\vskip.1in
\noindent
Work of L.Sz. is supported by the Polish Grant 1 P03B 028 28. 
He is a Visiting Fellow of the FNRS (Belgium). I.V.A. thanks NATO for a grant.

\end{document}